\documentclass[floats,floatfix,showpacs,amssymb,prd,twocolumn,superscriptaddress,nofootinbib,nolongbibliography,reprint]{revtex4-1}

\usepackage{amssymb,amsmath,verbatim,mathtools,needspace,enumitem,etoolbox,graphicx,physics,microtype,afterpage,bigints,gensymb,tabularx}

\usepackage[dvipsnames, usenames]{xcolor}
\definecolor{linkcolor}{rgb}{0.0,0.3,0.5}
\definecolor{dodgerblue}{HTML}{1E90FF}
\usepackage[unicode, colorlinks=true, linkcolor=linkcolor, citecolor=linkcolor, filecolor=linkcolor,urlcolor=linkcolor, pdfusetitle]{hyperref}
\usepackage[all]{hypcap}
\usepackage[T1]{fontenc}
\usepackage[utf8]{inputenc}
\usepackage{orcidlink}
\usepackage[htt]{hyphenat}
\usepackage{graphicx}

\usepackage{rotating}
\usepackage{tikz}

\newcommand{\ssim}{\mathchar"5218\relax\,}

\makeatletter
\newcommand*{\balancecolsandclearpage}{\close@column@grid \cleardoublepage \twocolumngrid}	
\makeatother

\usepackage{lipsum}

\newcommand{\milan}{\affiliation{Dipartimento di Fisica ``G. Occhialini'', Universit\'a degli Studi di Milano-Bicocca, Piazza della Scienza 3, 20126 Milano, Italy}}
\newcommand{\infn}{\affiliation{INFN, Sezione di Milano-Bicocca, Piazza della Scienza 3, 20126 Milano, Italy}}
\newcommand{\mrs}{\affiliation{Aix-Marseille Universit\'e, Universit\'e de Toulon, CNRS, CPT, Marseille, France}}

\begin{document}

\title{Sampling the full hierarchical population posterior distribution
\\  in gravitational-wave astronomy
}

\author{Michele Mancarella$\,$\orcidlink{0000-0002-0675-508X}}
\email{mancarella@cpt.univ-mrs.fr}
\mrs

\author{Davide Gerosa$\,$\orcidlink{0000-0002-0933-3579}}

\milan \infn


\date{\today}

\begin{abstract}

We present a full sampling of the hierarchical population posterior distribution of merging black holes using current gravitational-wave data. We directly tackle the most relevant intrinsic parameter space made of the binary parameters (masses, spin magnitudes, spin directions, redshift) of all the events entering the GWTC-3 LIGO/Virgo/KAGRA catalog, as well as the hyperparameters of the underlying population of sources. This results in a parameter space of about 500 dimensions, in contrast with current investigations where the targeted dimensionality is drastically reduced by marginalizing over all single-event parameters. 
In particular, we have direct access to (i) population parameters, (ii) population-informed single-event parameters, and (iii) correlations between these two sets of parameters. 
We quantify the fractional contribution of each event to the constraints on the population hyperparameters.
Our implementation relies on modern probabilistic programming languages and Hamiltonian Monte Carlo, with a continuous interpolation of single-event posterior probabilities. Sampling the full hierarchical problem is feasible, as demonstrated here, and advantageous as it removes some (but not all) of the Monte Carlo integrations that enter the likelihood together with the related variances. 
\end{abstract}

\maketitle

\section{Introduction}

Gravitational-wave (GW) astronomy employs hierarchical Bayesian inference~\cite{Mandel:2018mve,Vitale:2020aaz} to determine the population properties of merging compact binaries. After individual events are detected and characterized from the strain data~\cite{LIGOScientific:2018mvr,LIGOScientific:2020ibl,LIGOScientific:2021usb,KAGRA:2021vkt}, the posterior distribution on their parameters serve as the starting point for population analyses, where the ensemble of these events is analyzed by propagating the individual uncertainties and correcting for selection effects~\cite{KAGRA:2021duu}.
 
These analyses target parameters that describe the astrophysical population from which the events are drawn, such as mass cutoffs~\cite{Talbot:2018cva}, spectral indices, and parameters modeling the redshift evolution of the merger rate~\cite{Fishbach:2017zga,Fishbach:2018edt} or the spin distribution~\cite{Talbot:2017yur,Wysocki:2018mpo} (see~\cite{Callister:2024cdx} for a recent review). These can be, in turn, related to the formation and evolution of compact objects \cite{Mandel:2018hfr,Mapelli:2021taw}.
As the number of detected sources increases, more complex questions can now be posed to the data. Currently, the focus is shifting from marginal population distributions to exploring correlations between parameters at the population level~\cite{Callister:2021fpo,Biscoveanu:2022qac,Adamcewicz:2022hce,Adamcewicz:2023mov,Heinzel:2023hlb,Rinaldi:2023bbd,Pierra:2024fbl,Heinzel:2024hva,Lalleman:2025xcs}.
 Investigating more complex models and larger datasets requires greater numerical accuracy~\cite{Farr:2019rap,Essick:2022ojx,Talbot:2023pex}.

In its most common implementation, population inference estimates hyperparameters while marginalizing over individual-event parameters at evaluation time.
While this approach reduces the dimensionality of the problem, it comes at the cost of evaluating computationally expensive and potentially inaccurate Monte Carlo integrals, whose limited accuracy may restrict the range of accessible models.
This issue is linked to a more fundamental point: at the statistical level, individual event and population parameters are part of the same, full, high-dimensional hierarchical Bayesian model, with their interplay providing valuable physical insights. In particular, knowledge of the population properties can inform the individual properties of each event in the sample. Hierarchical inference provides a ``population-informed prior'' for individual events that may change, or at least influence, our physical interpretation~\cite{Moore:2021xhn}. Similarly, correlations between individual events and population parameters can help identify which events are most informative about specific features, such as gaps and edges in the mass spectrum~\cite{Essick:2021vlx}. 
Currently, these correlations are lost in the marginalization of the hierarchical posterior and have to be obtained in post-processing through non-trivial reweighting schemes~\cite{Moore:2021xhn, DCC_Fishbach,DCC_Callister}.

This paper tackles the full hierarchical problem for current detections of merging black holes (BHs). First and foremost, we demonstrate that high-dimensional inference of the complete parameter space is feasible. Our implementation leverages modern probabilistic programming languages~\cite{rainforth2017automating} (PPLs), which have significantly advanced Bayesian modeling by providing flexible, scalable, and computationally efficient tools for complex hierarchical models. PPLs such as \textsc{PyMC}~\cite{Salvatier2016}, \textsc{Stan}~\cite{Carpenter2017}, and \textsc{NumPyro}~\cite{phan2019composable,bingham2019pyro} allow specifying probabilistic models using intuitive, high-level syntax. Furthermore, by leveraging automatic differentiation~\cite{Baydin2018}, they facilitate the optimization of complex likelihood functions via gradient-based sampling algorithms such as Hamiltonian Monte Carlo (HMC)~\cite{Duane1987,Brooks_2011} and Variational Inference~\cite{Blei2017}. 
These languages are increasingly employed in GW astronomy, see e.g.~Refs.~\cite{KAGRA:2021duu,Callister:2022qwb,Chen:2024gdn,Leyde:2024tov}.
Here, we make use of PPLs and HMC to sample the full population posterior distribution.
This provides direct access to single-event parameters, population parameters, and their correlations while mitigating some numerical challenges in the likelihood evaluation due to the drastic reduction of the Monte Carlo integrals involved.
Similar approaches have so far occasionally been applied to simulated data~\cite{Farr:2019twy,Mancarella:2024qle} (also see~Ref.~\cite{Mould:2023eca} for a one-dimensional application to GWTC-3 data) and are posed to become increasingly relevant with the large GW catalogs expected in the near future~\cite{Baibhav:2019gxm,KAGRA:2013rdx}.
We provide an open--source implementation at Ref.~\cite{pymcpop}

\section{Methods}\label{sec:methods}

\subsection{Population inference}

Consider a detection catalog of $N_{\rm obs}$ events labeled by~$i$. We denote the stretch of data related to each event by $\mathcal{D}_i$ and the full dataset entering the catalog by~$\{\mathcal{D}\}$. The implicit assumption here is that events are statistically independent and do not overlap in the same stretch.

Each event is described by parameters $ {\theta}$;  the cardinality of this parameter set is denoted by $N_\theta$. The set $\{ \theta\}$ denotes all the single-event parameters of all the events and has cardinality $N_\theta\times N_{\rm obs}$. 
We put ourselves in the situation where each GW event has already been analyzed assuming a certain ``parameter-estimation'' prior $\pi_{\rm PE} (\theta_i)$ and we have access to the posterior $p(\theta_i | \mathcal{D}_i)$. This is the case when using public data products from the LVK Collaboration. 

Much like in most of the current GW population analyses, we consider the following $N_\theta = 7$ parameters
 \begin{align}\label{thetapar}
 \theta = \{  m_1, m_2, z, \chi_1, \chi_2, \cos \theta_1, \cos \theta_2  \}\,,
 \end{align}
  where $m_{1}>m_{2}$ are the source-frame masses, $z$  is the redshift, $\chi_{1,2}$ are the spin magnitudes, and $\cos \theta_{1,2}$ are the spin tilts. 
The event parameters $\theta$ are assumed to be drawn from a population distribution $p_{\rm pop}(\theta | \lambda)$, which in turn depends on a set of parameter $\lambda$ with cardinality $N_{\lambda}$.  This number is of $\mathcal{O}(10)$ for state-of-the-art applications, and more specifically $N_{\lambda}=13$ for the population model adopted in this paper.
Note that the individual-event posteriors depend on more parameters than those listed in Eq.~(\ref{thetapar}) (e.g., sky location, polarization, etc.): by omitting these additional quantities, we are implicitly assuming that their astrophysical distribution $p_{\rm pop}$ coincides with~$\pi_{\rm PE}$. 

The hierarchical posterior distribution of an 
inhomogeneous Poisson process in presence of selection effects is given by \cite{Loredo:2004nn,Mandel:2018mve,Vitale:2020aaz}
\begin{equation}\label{h_post}
  p\big(  {\lambda}, \{  {\theta}\} | \{ {\cal D} \}  \big) \propto 
    \frac{\pi( {\lambda})}{ \xi( {\lambda})^{N_{\rm obs}}}
    \prod_{i=1}^{N_{\rm obs}} p\big( {\theta}_i   |  {\cal D}_i \big) \frac{p_{\rm pop}( {\theta}_i |  {\lambda}) }{\pi_{\rm PE}( {\theta}_i)}   
       \, , 
       \end{equation}
where  $\pi(\lambda)$ is a prior distribution. In practice, the term inside the product sign reweights the event posterior $p\big( {\theta}_i   |  {\cal D}_i \big)$ from the default prior ${\pi_{\rm PE}( {\theta}_i)} $ to the population prior $p_{\rm pop}( {\theta}_i |  {\lambda})$. Equation (\ref{h_post}) further assumes that the parameters $\lambda$ carry information solely about the shape of the BH-binary distribution and that the event rate $R$ follows a scale-invariant prior $\propto 1/R$. 

Crucially, the probability density $p_{\rm pop}(\theta|\lambda)$ models the intrinsic (i.e. astrophysical) distribution of sources, but our detectors observe a small fraction of it. Selection effects are captured by the term $ \xi( {\lambda})$, which is the fraction of detected events in the targeted population: 
\begin{equation}\label{xi}
\xi({\lambda}) = \int \! \dd \theta \,  p({\rm det} | {\theta}) \,   p_{\rm pop}( {\theta} |  {\lambda}) \, ,
\end{equation}
where $p({\rm det} | {\theta}) \in [0,1]$ is the probability of detecting an event with parameters  $\theta$. For a faithful reconstruction~\cite{2019PhRvD.100h3015W}, the adopted model of $p({\rm det} | {\theta})$ must match the detection strategy that populated the GW catalog in the first place. This implies that the detectability depends on the data and not on the source parameters present in those data~\cite{Essick:2023upv}. 
In current GW-astronomy applications, the detection efficiency $\xi({\lambda})$ is estimated by re-weighted Monte Carlo integration~\cite{Tiwari:2017ndi,Farr:2019rap} from a set of injections in the detection pipelines. In practice, one performs $N_{\rm draw}$ injections with parameters $\theta_k$ from a nominal population $p_{\rm draw}({\theta}_k)$ and applies the same thresholding criterion that was used to select the event entering the detection catalog (for instance, Ref.~\cite{KAGRA:2021duu} considered search triggers with detection false-alarm-rate $< 1 \, {\rm yr}^{-1}$). This means  that $p({\rm det} | \theta_k) = 1$ for the $N_{\rm det}$ injections that pass the selection threshold and $p({\rm det} | \theta_k)= 0$ otherwise. The Monte-Carlo estimator of $\xi({\lambda})$ is
\begin{equation}\label{xihat}
\hat \xi( {\lambda})  = \frac{1}{N_{\rm draw}}\, \sum_{k=1}^{N_{\rm det}} \frac{p_{\rm pop}({\theta}_k |  {\lambda}) }{p_{\rm draw}({\theta}_k)} \,.
\end{equation}

Equation~(\ref{h_post}) is a probability density over $N_{\lambda} + N_{\rm obs} \times N_{ {\theta}}$ dimensions. For the current GW dataset and targeted models, this number is $\ssim 500$. To simplify the problem, one can marginalize over all the individual event parameters and consider 
\begin{equation}\label{h_post_marginal}
  p\big(  {\lambda} | \{ {\cal D} \}  \big) \propto 
    \frac{\pi( {\lambda})}{ \xi( {\lambda})^{N_{\rm obs}}}
    \prod_{i=1}^{N_{\rm obs}} \int \! \dd  {\theta}_i \,   \frac{p\big( {\theta}_i   |  {\cal D}_i \big)}{\pi_{\rm PE}( {\theta}_i)}   
     p_{\rm pop}( {\theta}_i |  {\lambda})  \, .
\end{equation}
While this is a much simpler, $N_{\lambda}$-dimensional problem, one needs to compute $N_{\rm obs}$ integrals in $N_{ {\theta}}$ dimensions. These are typically approximated via Monte Carlo using samples  $\theta_{k,i}\sim p\big( {\theta}_i   |  {\cal D}_i \big)$ from the individual-event posteriors. The corresponding estimator is \begin{equation}\label{Lambdaihat}
\hat {\mathcal{L}}_i   = \frac{1}{{N_{{\rm s}, i}}}\, \sum_{k=1}^{N_{{\rm s}, i}} \frac{ p_{\rm pop}( {\theta}_{k,i} |  {\lambda})}{\pi_{\rm PE}( {\theta}_{k, i})} \,.
\end{equation}
In practice, this allows for an efficient ``recycling'' of the single-event analyses into population studies. 

\subsection{Sampling strategy}

Instead of marginalizing over the event parameters, the goal of this paper is to sample the full posterior of Eq.~(\ref{h_post}).
The approach that most closely follows the logic of a hierarchical model would be to write the posterior as 
\begin{align}
  p\big(  {\lambda}, \{  {\theta}\} | \{ {\cal D} \}  \big) & \propto 
    {\pi( {\lambda})}  \left [  \prod_{i=1}^{N_{\rm obs}}  {p_{\rm pop}( {\theta}_i |  {\lambda}) } \right]  
  \notag  \\
    &\times
    \left [ \prod_{i=1}^{N_{\rm obs}} \frac{p\big( {\theta}_i   |  {\cal D}_i \big) }{\pi_{\rm PE}( {\theta}_i)}   \right]  
     { \xi( {\lambda})^{- N_{\rm obs}}}
       \, , 
\end{align}
and proceed with the following implementation:
\begin{enumerate}[label=(\roman*)]
\item Sample $ {\lambda} \sim \pi( {\lambda})$;
\item Sample $ {\theta}_i \sim p_{\rm pop}( {\theta}_i |  {\lambda})$ for $i=1,\dots, N_{\rm obs}$; 
\item Evaluate the product 
$\prod_{i=1}^{N_{\rm obs}} p\big( {\theta}_i   |  {\cal D}_i \big) / \pi_{\rm PE}( {\theta}_i)  $;
\item Compute $\xi(\lambda)$. 
\end{enumerate}
However, this can be computationally inefficient whenever the population prior is much broader than the individual-event likelihoods (which is generally the case, even for current GW catalogs, at least for binary BHs). 

We will instead follow a different approach and write
\begin{align}\label{h_lik_simple}
  p\big(  {\lambda}, \{  {\theta}\} | \{ {\cal D} \}  \big) & \propto 
    {\pi( {\lambda})}  \left [  \prod_{i=1}^{N_{\rm obs}}  { p\big( {\theta}_i   |  {\cal D}_i \big) } \right]  
  \notag  \\
    &\times
    \left [ \prod_{i=1}^{N_{\rm obs}} \frac{
    p_{\rm pop}( {\theta}_i |  {\lambda})
    }{\pi_{\rm PE}( {\theta}_i)}   \right]  
     { \xi( {\lambda})^{- N_{\rm obs}}}
       \, . 
\end{align}
While this is a trivial manipulation of the previous expressions, it highlights a different implementation strategy:
\begin{enumerate}[label=(\roman*)]
\item Sample $ {\lambda} \sim \pi( {\lambda})$;
\item Sample $ {\theta}_i \sim p\big( {\theta}_i   |  {\cal D}_i \big)$ for $i=1,\dots, N_{\rm obs}$; 
\item Evaluate the product  
$\prod_{i=1}^{N_{\rm obs}} p_{\rm pop}( {\theta}_i |  {\lambda}) / \pi_{\rm PE}( {\theta}_i)  $;
\item Compute $\xi(\lambda)$.
\end{enumerate}
When using existing, pre-packaged stochastic samplers, the procedure above can be easily achieved by implementing  $  \pi_{  {\cal D}}( \{  {\theta} \})  \equiv  \prod_{i=1}^{N_{\rm obs}}  p\big( {\theta}_i   |  {\cal D}_i \big)$ as a prior distribution on the single event parameters. ``Prior'' here is an improper word because this term depends on the data, but it can be implemented as a prior nonetheless. This is because, once more, we wish to recycle existing posterior samples: the single-event parameter-estimation runs are performed \emph{prior to} and independently of the population analysis.

This approach requires a continuous interpolation of the single-event posterior distributions $p\big( {\theta}_i   |  {\cal D}_i \big)$, both to ensure differentiability and to avoid limitations due to the finite number of available posterior samples.
Rather than working with the variables $ {\theta}$ of Eq.~(\ref{thetapar}), we switch to
\begin{align}\label{theta_tilde}
\tilde{ {\theta}} = & \Bigg\{  \log \mathcal{M}_{z}, \log \frac{q}{1-q},  \log d_L,    \log \frac{\chi_1}{1-\chi_1} , \notag \\
& \log \frac{\chi_2}{1-\chi_2}, \log \frac{1+\cos \theta_1}{1-\cos \theta_1} , \log \frac{1+\cos \theta_2}{1-\cos \theta_2}  \Bigg\}\,,
\end{align}
where $ \mathcal{M}_{z} = (m_1 m_2)^{3/5}  (1+z)/ (m_1+m_2)^{1/5} $ is the detector-frame chirp mass, $q = m_2/m_1$ is the mass ratio, and $d_L$ is the luminosity distance. 
The motivation behind this choice is twofold. First, detector-frame quantities are obtained directly from publicly available single-event posterior samples without a cosmology-dependent transformation; second, this maps bounded variables to unbounded domains, thus avoiding the presence of sharp edges when interpolating.\footnote{Specifically, masses and distances are defined to be positive, with the secondary mass further obeying the constraint $m_2 \leq m_1$. %, i.e. $q \leq 1$.  
 Spins have magnitudes $\in[ 0, 1]$ and cosines of tilt angles  $\in[-1,1]$.} 
With the variables $\tilde{ {\theta}}$, the only difference in the sampling procedure above is that step (ii) is modified into sampling $ \tilde{\theta}_i \sim p\big( \tilde{\theta}_i   |  {\cal D}_i \big)$ and inverting Eq.~(\ref{theta_tilde}) to compute $ {\theta}(\tilde{\theta}_i)$. The variables $ {\theta}$ obtained this way are draws from the distribution $p\big( {\theta}_i   |  {\cal D}_i \big) = p\big( \tilde{\theta}_i   |  {\cal D}_i \big) \, | \dd  \tilde{\theta}_i/ \dd {\theta}_i |$.

We interpolate the individual posterior distributions $ p(\tilde{ {\theta}}_{i} |  \{ {\cal D}_i  \} ) $ using a Gaussian Mixture Model as implemented in {\sc scikit-learn}~\cite{pedregosa2011scikit}. The number of components is chosen to minimize the Bayesian Information Criterion (BIC) \cite{10.1214/aos/1176344136} to avoid overfitting. 
In particular, once a minimum is found, we check that increasing the number of components by at least 10 does not improve the BIC further.
We find that the BIC prefers a number of components between 5 and 257 for the various events in the GW catalog.

As for the sampler, we use {\sc PyMC}~\cite{pymc2023}, an open-source probabilistic programming library in {\sc Python} that leverages {\sc PyTensor} \cite{pytensor} (formerly {\sc Theano} and {\sc Aesara}) as its computational engine while also allowing compilation on popular backends such as {\sc JAX}~\cite{Bradbury2018} and {\sc Numba}~\cite{Lam2015}.
In particular, we use No U-Turn Sampling (NUTS) \cite{JMLR:v15:hoffman14a}, which is a flavor of HMC~\cite{Duane1987,Brooks_2011}.

\subsection{Accuracy requirements} \label{accreq}

Whenever the likelihood evaluation involves Monte Carlo integrals, one must be careful with the additional uncertainty introduced by this technique. This is a pressing issue in GW astronomy and mitigation strategies are under active development~\cite{Farr:2019rap,Essick:2022ojx,Talbot:2023pex}.

For a given probability distribution function $p(x)$ with $\int \dd x \; p(x)=1$, the expectation value of a generic quantity $f(x)$ is given by \begin{equation}
\langle f \rangle =   \int \! \, \dd x \, f(x) p(x)\,.
\end{equation}
Assuming draws $x_k\sim p(x)$ with $k=1,\dots, N_{\rm draw}$, the Monte Carlo estimator of $\langle f \rangle $ is the arithmetic mean
\begin{equation}
\hat{\langle f \rangle} =  \frac{1}{N_{\rm draw}} \, \sum_{k=1}^{N_{\rm draw}} f(x_k) \, 
\end{equation}
and the variance on this estimate is given by 
\begin{equation}
{\rm var} \langle f \rangle = \frac{1}{N_{\rm draw}} \, \left[ \langle f^2\rangle - \langle f\rangle ^2 \right]\,.
\end{equation}

Intuitively, one needs a large enough number of samples $N_{\rm draw}$ such that this variance is sufficiently small.
One possibility is to impose a condition on the effective number of independent samples~\cite{Farr:2019rap}
\begin{equation}\label{Neff}
N_{\rm eff} \equiv  N_{\rm draw} \, \frac{\langle f\rangle ^2}{ \langle f^2\rangle}\,. 
\end{equation}
More specifically, a number of effective samples $N_{\rm eff} >  4\, N_{\rm obs}$  is often used 
when evaluating the selection function $\xi(\lambda)$, with the interpretation that otherwise the injection dataset would not sufficiently cover the targeted population~\cite{Farr:2019rap}. For the integrals appearing at the numerator of the marginalized likelihood of Eq. (\ref{h_post_marginal}), previous studies~\cite{KAGRA:2021duu} have enforced $N_{{\rm eff} , i}>  N_{\rm obs}$ for each term in the product.  
More recently, Ref.~\cite{Talbot:2023pex} pointed out that it might be preferable to work directly with the variance for the estimator of the full population log-likelihood, ${\rm var}(\log \mathcal{L})$. For the marginalized likelihood of Eq.~(\ref{h_post_marginal}), this is obtained by error propagation as [cf.  Eqs.~(\ref{xihat}-\ref{Lambdaihat})]
\begin{equation}
{\rm var}(\log \mathcal{L}) =  \sum_{i=1}^{N_{\rm obs}} \frac{{\rm var}\mathcal{L}_i }{ \hat{\mathcal{L}}_i^2} + N_{\rm obs}^2 \, \frac{{\rm var} \,\xi }{ \hat{\xi}^2} \, .
\end{equation}
Reference~\cite{Talbot:2023pex} finds that a threshold ${\rm var}(\log \mathcal{L})<1$ is necessary to ensure the reliability of the final population posterior distribution.  

In practice, these conditions are used as regularization strategies: the sampler is explicitly forbidden to explore regions of the parameter space where the Monte Carlo variances are too large. 
Addressing which regions of the parameter space are affected by these cuts is challenging, as this depends on both the estimate of the selection function (which in turn depends on $\lambda$) and the individual-event posterior samples (which instead depend on each of the $\theta_i$). 
These conditions are data-dependent, so their implementation corresponds to modifying the likelihood. In practice, however, they are not intrinsic to the likelihood itself, but originate from numerical and technical limitations.

In the case of this paper, sampling the full posterior distribution over both $\lambda$ and $\{\theta\}$ eliminates the need for Monte-Carlo integrating the single-event likelihoods. The price to pay is that of a tougher requirement on the convergence of the sampler, which now has to explore a larger parameter space, but, crucially, there is no variance associated with this at each iteration.\footnote{Note, however, that there can be a small systematic uncertainty due to the continuous interpolation of the posterior samples, which should be kept under control. However, the posterior samples are themselves uncertain because of their finite number.} We are then left with 
\begin{equation}
{\rm var}(\log \mathcal{L}) = N_{\rm obs}^2 \, \frac{{\rm var} \, \xi }{ \hat{\xi}^2} =  \frac{ N_{\rm obs}^2} { N_{\rm eff}} \left[ 1- \frac{N_{\rm eff}}{N_{\rm draw}} \right]\, .
\end{equation}
In the limit where the number of pipeline injections  $N_{\rm draw}$ is large, one has ${\rm var}(\log \mathcal{L}) \simeq { N_{\rm obs}^2}/{ N_{\rm eff}}$. The condition ${\rm var}(\log \mathcal{L}) < 1$ thus translates to $N_{\rm eff} >  N_{\rm obs}^2$, which is a much more stringent condition on the $\xi(\lambda)$ integral compared to the usually adopted threshold $N_{\rm eff} > 4\, N_{\rm obs}$. 
We adopt the condition  ${\rm var}(\log \mathcal{L})<1$  throughout the paper and discuss the impact of this choice in Appendix~\ref{sec:accuracy}.

\begin{figure*}[t]
  \includegraphics[width=0.96\textwidth]{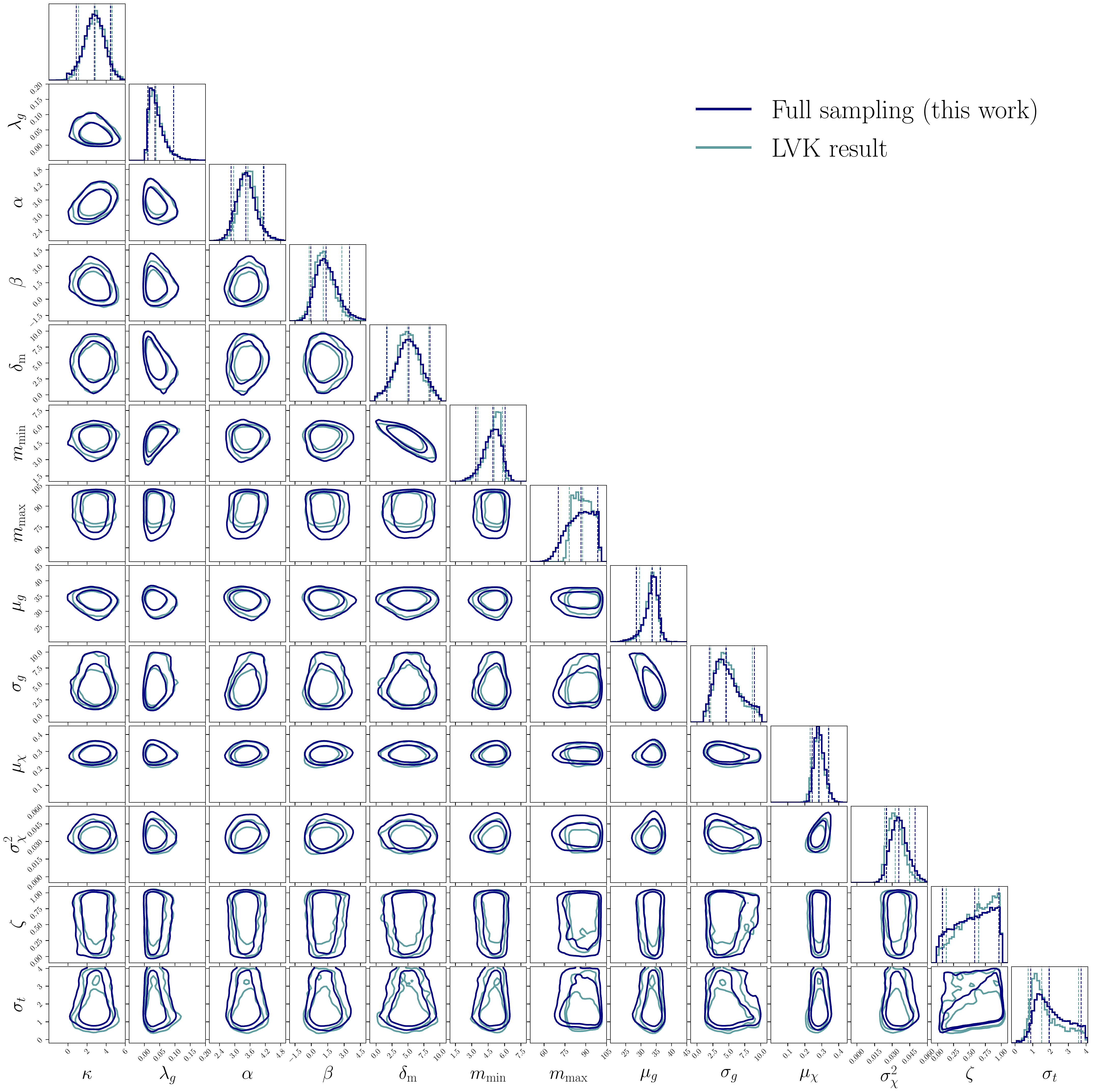}
    \caption{ 
    Posterior distribution of the population hyperparameters, assuming a fixed cosmology. The dark blue distribution refers to our full hierarchical sampling of Eq.~(\ref{h_post}). The light blue distribution reports results from Ref.~\cite{ KAGRA:2021duu}, which are restricted to the marginalized posterior of Eq.~(\ref{h_post_marginal}). Contours correspond to $68\%$ and $90\%$ credible intervals.
}   \label{fig:corner_with_LVK}
\end{figure*}

\subsection{Population model and data}

\begin{figure*}[th]
    \includegraphics[width=\textwidth]{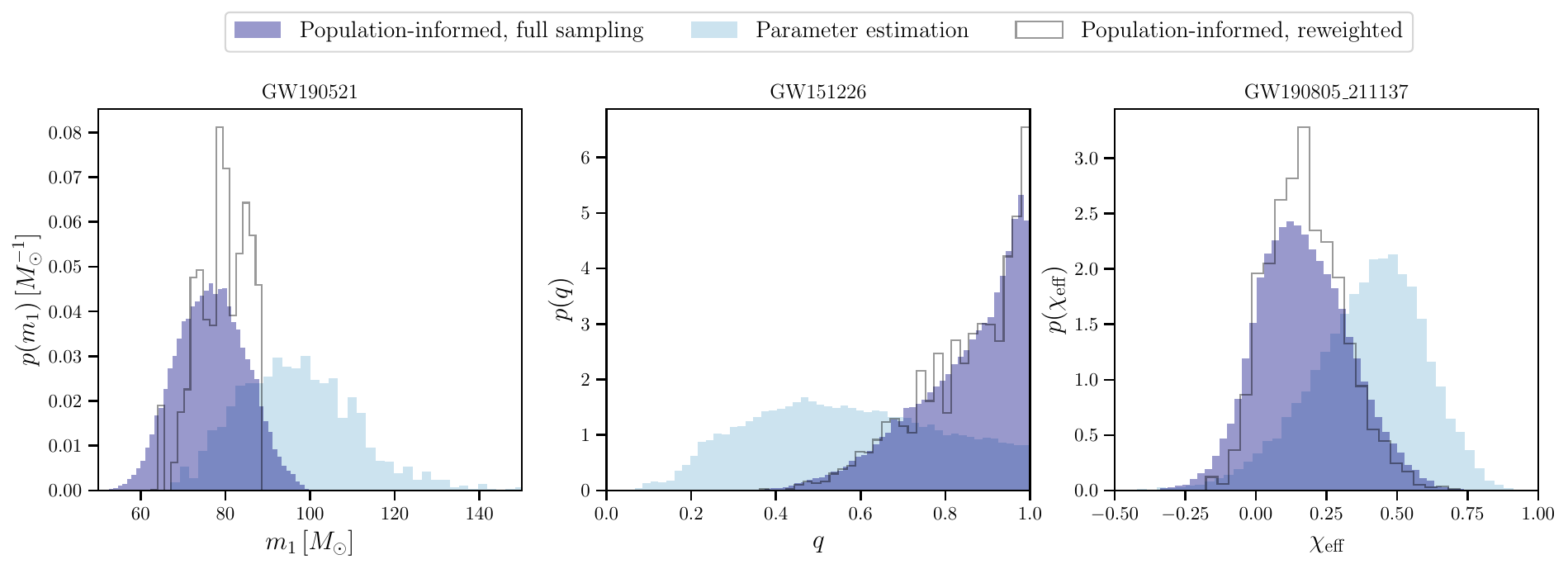}
    \caption{Single-event marginal posterior distributions for some representative events and some representative parameters. We show results from our full inference run (dark blue, filled), parameter-estimation results obtained with uninformative priors (light blue, filled), and population-informed reweighted results from Ref.~\cite{KAGRA:2021duu} (black, empty)
   }
 \label{fig:marginal_m1_190521}
\end{figure*}

We adopt a standard parametrization of the binary BH population $p_{\rm pop}(\theta | \lambda)$ to compare against Ref.~\cite{KAGRA:2021duu}. The population prior is factored as 
\begin{align}
p_{\rm pop}(\theta|\lambda)& = p(m_1,m_2 | \lambda_m)\, p(\chi_1 | \lambda_\chi)\, p(\chi_2 | \lambda_\chi) 
\notag \\
& \times p(\theta_1,\theta_2 | \lambda_\theta)\, p(z|\lambda_z) \,,
\end{align}
where $\lambda = \{\lambda_m, \lambda_z, \lambda_\chi, \lambda_\theta\}$. 
We use the {\sc Power-Law + Peak} model for the mass distribution, which has eight free hyperparameters $\lambda_m=\{ \alpha_m, \beta_q, \lambda_{\rm peak}, \mu_m, \sigma_m, m_{\rm min}, m_{\rm max}, \delta_{\rm m} \}$, a power-law model for the redshift distribution, which has a single hyperparameter $\lambda_z = \{\kappa\}$, and LVK's {\sc Default} model for the spins where $\lambda_\chi =\{ \alpha_\chi, \beta_\chi\}$ and
$\lambda_\theta =\{ \zeta,\sigma_t\}$; see Ref.~\cite{KAGRA:2021duu} for details. 
In this case, the cosmology is fixed to that of Planck 2018~\cite{Planck:2018vyg}.

We also consider inference on the cosmological parameters within a flat $\Lambda$CDM scenario. In this case, the hyperparameters $ {\lambda}$ are augmented by  $ {\lambda}_{\rm c} = \{ H_0, \Omega_{\rm m} \}$ where  $H_0$ is the Hubble constant and $ \Omega_{\rm m}$ is the matter density. 
Note that, when cosmology is allowed to vary, the population prior in Eq.~(\ref{h_lik_simple}) inherits a dependence on the cosmological parameters when converting from the detector frame to the source frame. Additionally, cosmology enters a Jacobian factor to compute the single-event source--frame prior from the detector-frame one,  $\pi_{\rm PE}(\theta) =  \pi_{\rm PE, D}(\theta_{\rm D}) \, |\dd \theta_{\rm D}/\dd \theta| $, and through a comoving-volume factor contained in the redshift population prior (see e.g., Ref.~\cite{Mancarella:2021ecn} for a detailed discussion).

We use the same 69 BH binary events as in Ref.~\cite{KAGRA:2021duu}, which were detected with false-alarm rate $< 1\,{\rm yr}^{-1}$. 
We use public posterior samples from the GWTC-2.1 and GWTC-3 catalogs~\cite{KAGRA:2023pio, ligo_scientific_collaboration_and_virgo_2022_6513631, ligo_scientific_collaboration_and_virgo_2023_8177023}. 
Specifically, we selected samples flagged as \texttt{C01:IMRPhenomXPHM} and \texttt{*nocosmo.h5}.
Selection effects are computed as in Eq.~(\ref{xihat}) using public software injections into search pipelins~\cite{ligo_scientific_collaboration_and_virgo_2021_5636816}. Specifically, we use the \texttt{o1+o2+o3\_bbhpop\_real+semianalytic-LIGO-T2100377-v2}  injection set.
Priors on population hyperparameters match those of Ref.~\cite{KAGRA:2021duu}.

We run four independent chains of 20,000 samples each, after discarding an initial warm-up phase of 5,000 steps per chain, and use a target acceptance probability of 0.9.
We assess convergence by checking the Gelman-Rubin $\hat{r}$ statistics~\cite{10.1214/ss/1177011136}, finding $\hat{r}<1.01$ for all variables.
Our analysis took approximately ${\sim}50$ hours on four off-the-shelf CPUs. Note, however, the very large number of samples that were collected. We also stress that this implementation does not require access to an HPC facility. Further optimization is possible to improve performance, including execution on GPUs. The runtime depends on the target acceptance probability and the specific implementation details of the variance cuts.

\section{Validation}\label{sec:validation}

\subsection{Population parameters}

First, we compare our results (at fixed cosmology) with the standard population analyses that rely on the marginalized version of the posterior, here reported in  Eq.~(\ref{h_post_marginal}).
Figure~\ref{fig:corner_with_LVK} shows the posterior distribution in the 13-dimensional subspace of the population hyperparameters, obtained from the sampling presented here and the corresponding results from Ref.~\cite{KAGRA:2021duu}.

Considering our result is marginalized over $N_\theta \times N_{\rm obs} = 483$ dimensions, the agreement between the two distributions is excellent. Our full posterior sampling presents a slightly longer tail in the marginal posterior distribution of the parameter $m_{\rm max}$, which quantifies the upper cutoff of the primary mass spectrum. This population parameter is largely determined by the heaviest event in the population (c.f. Sec.~\ref{sec:novelty}); the difference reported here can be specifically attributed to the posterior distribution samples for the primary mass of $\rm GW190521\_030229$ (hereafter, and in all figure labels, simply 190521). Our interpolation combined with HMC leads to a smooth tail; on the other hand, when sampling the marginal likelihood, such a tail is dominated by the variance cut imposed on Monte Carlo integrals and results in a steeper decrease.

\subsection{Population-informed single-event parameters}

Sampling the full hierarchical problem provides population-informed posterior distributions of the individual event parameters \cite{Moore:2021xhn}. We compare our results against those of  Ref.~\cite{KAGRA:2021duu} where, crucially, their distributions have been obtained using a-posteriori reweighting schemes \cite{DCC_Fishbach,DCC_Callister}. 

Figure~\ref{fig:marginal_m1_190521} shows the marginal posterior probabilities of three illustrative cases, namely, the most massive event in the catalog (GW190521) and its primary mass $m_1$ (left), an event that presents extensive support for small mass ratio $q$ when analyzed using uninformative priors (GW151226) (center), and an event with support for non-zero effective spin $\chi_{\rm eff}$ ($\rm GW190805\_ 211137$) (right).
Results from our inference run are compared against samples obtained under the uninformative ``parameter estimation'' prior, as well as the population-informed results reconstructed via reweighting.
Our implementation leads to very compatible results, crucially with much better sampling coverage. Reweighting schemes inevitably lead to undersampling in cases where the base and the target distributions differ, but these are precisely the most interesting case where population-informed constraints can provide new insights.

\section{Novelty}\label{sec:novelty}

\begin{figure*}[th]
     \includegraphics[width=\textwidth]{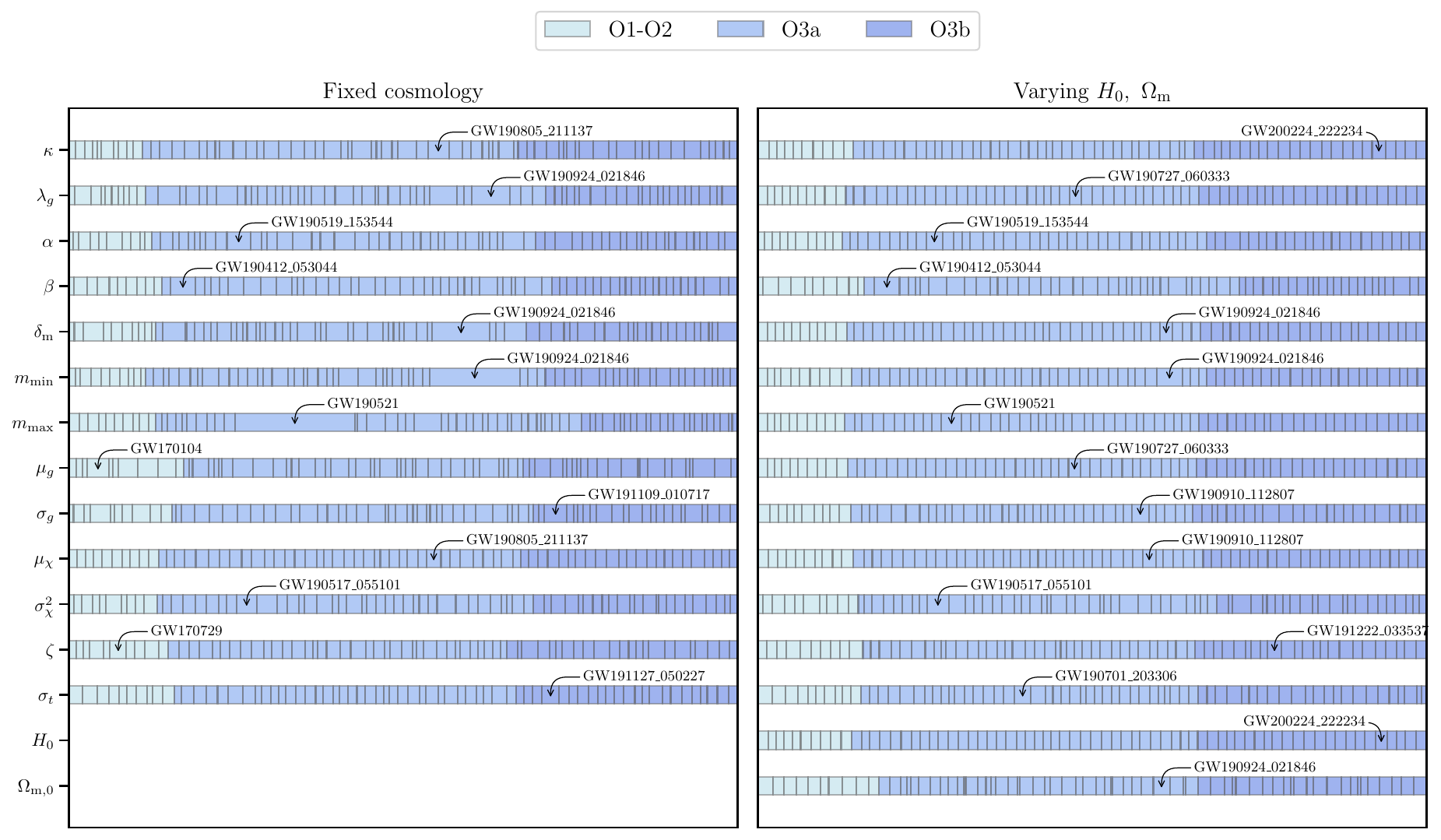}
    \caption{Fractional contribution from each event in the GWTC-3 catalog to the population hyperparameters constraints. This is quantified using the averaged correlation coefficient of Eq.~(\ref{omegaik}). Rows indicate the different hyperameters, either fixing (left) or varying (right) the cosmology. Vertical black lines indicate the fractional contribution of a given event in chronological order. Colors refer to the LVK data-taking periods.  For each hyperparameter, the event contributing the most is indicated explicitly. 
    }
    \label{fig:contributions}
\end{figure*}

We now discuss new possibilities opened by the full posterior sampling presented in this work and present a systematic investigation of correlations between population features and individual events. 

\subsection{Correlation overview}

The impact of specific events on the inferred population features is usually investigated in a targeted way, e.g., with leave-one-out strategies, examining the impact of the presence/absence of a specific event, or with posterior-predictive checks~\cite{KAGRA:2021duu,Essick:2021vlx,Fishbach:2019ckx,Miller:2024sui}. 
On the other hand, our full-inference approach allows us to directly inspect the contribution of each event in the catalog to any population feature, identifying ``special'' events and pinpointing hyperparameters that are particularly sensitive to them. 

To this end, we denote by $\rho_{ijk}$ the Pearson correlation coefficient between the single-event parameter $j$ for event $i$ and the population-level hyperparameter $k$. We then define
\begin{align}
\omega_{ik} = \frac{  \frac{1}{N_{\theta} }\sum_{j} | \rho |_{ijk}  }{ \sum_i \left(   \frac{1}{N_{\theta} }\sum_{j} | \rho |_{ijk}  \right) },
\label{omegaik}
\end{align}
which is the average, normalized correlation of each population hyperparameter with the parameters of each individual event, $\theta$. %

Figure~\ref{fig:contributions} shows the quantities $\omega_{ik}$ for each population-level hyperparameter, assuming both a fixed cosmology and varying $H_0, , \Omega_{\rm m}$. The segments in each of the corresponding horizontal bars are proportional to the contribution of each event $i$.
When fixing the cosmology, some population parameters show evenly spaced bars (indicating comparable contributions from all events), while others exhibit more structure. In the latter case, observing the right event is crucial, whereas in the former, collecting as many events as possible is more important. Notably, parameters defining the mass spectrum edges are primarily constrained by the heaviest (GW190521) and lightest ($\rm GW190924\_021846$) events.
On the other hand, constraints on the spin hyperparameters do not appear to be dominated by any individual event.
When, instead, the cosmological parameters are left to vary, we find a more equal contribution from all the events in the catalog. This supports the intuition that the cosmological constraints are driven by a coherent shift with redshift across the mass spectrum rather than by specific events~\cite{Ezquiaga:2022zkx}. 

 \begin{figure}
    \includegraphics[width=0.355\textwidth]{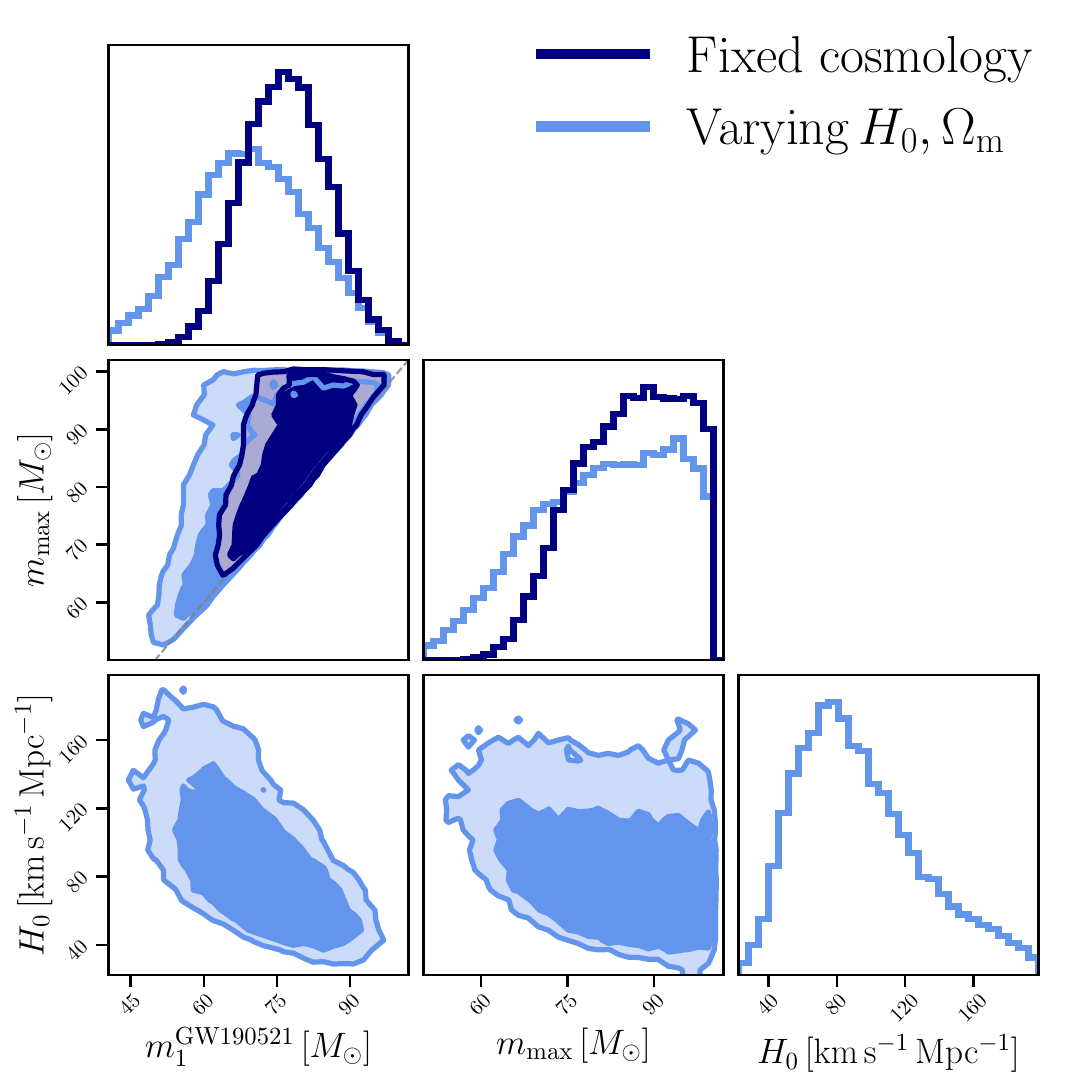}
  %  \hspace{2cm}
    \includegraphics[width=0.355\textwidth]{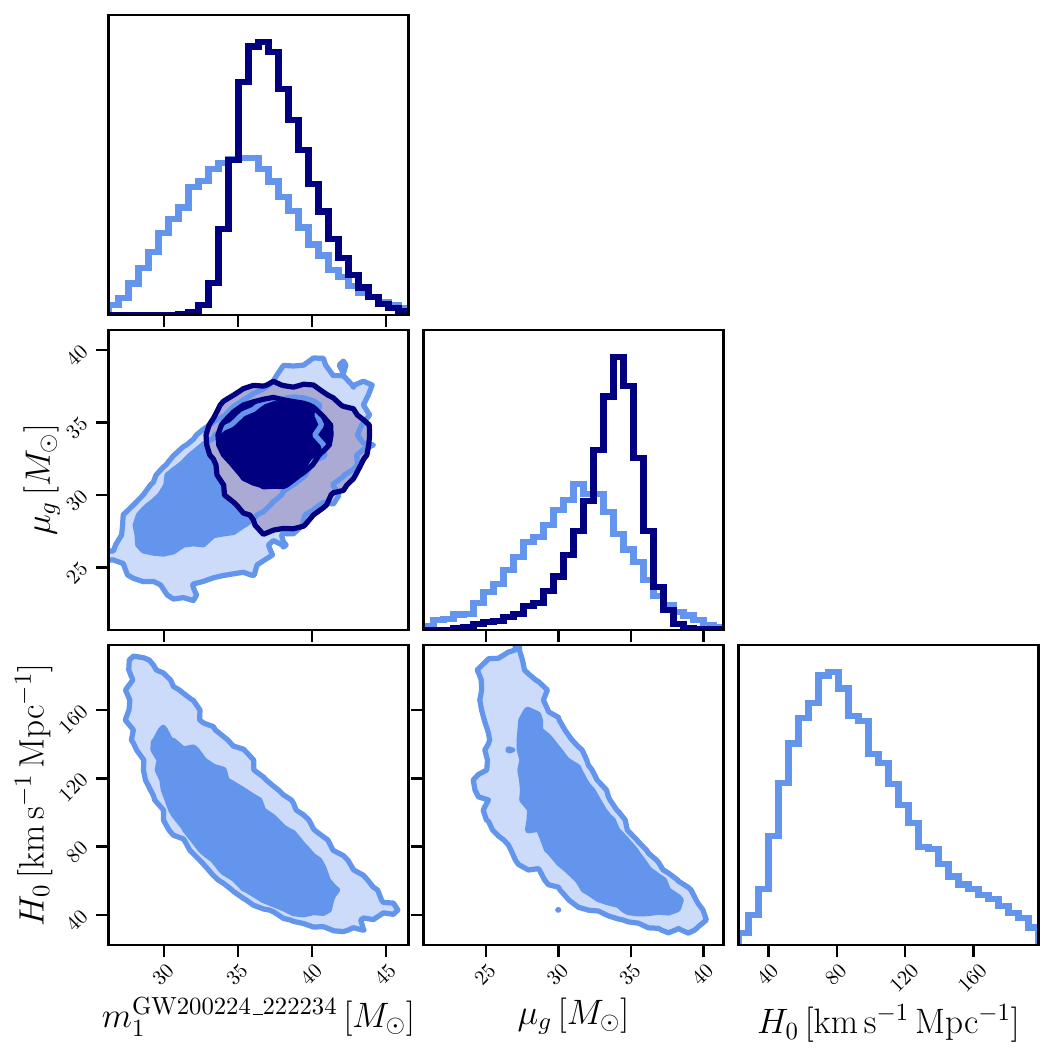}
        \includegraphics[width=0.355\textwidth]{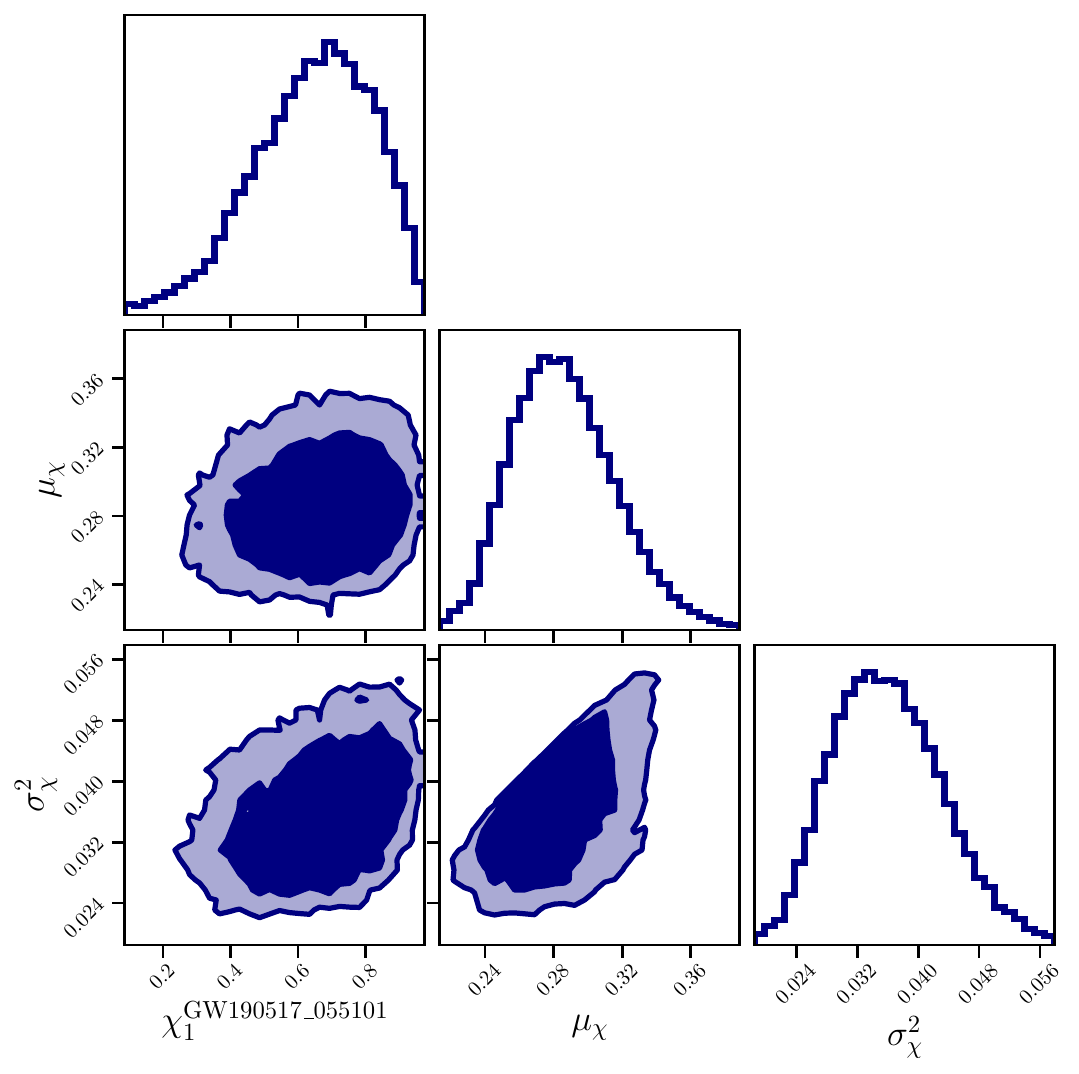}

    \caption{Examples of correlations between single-event parameters and population/cosmological parameters. The top panel shows the primary mass of GW190521, the upper mass cutoff, and the Hubble constant. The middle panel shows the primary mass of $\rm GW200224 \_222234$, the location of the Gaussian peak in the mass spectrum, and the Hubble constant. The bottom panel shows the primary spin of $\rm GW190727\_06033$, and the moments of the population spin distribution.  
    Dark blue (light blue) distributions refer to runs where we fix (vary) the cosmological parameters. Contours correspond to $68\%$ and $95\%$ credible intervals.}
    \label{fig:mass_correlations}
\end{figure}

\subsection{Correlation examples}

In Fig.~\ref{fig:mass_correlations}, we inspect more closely the correlation of single-event to population parameters for those events that mostly inform specific features (labeled explicitly in Fig.~\ref{fig:contributions}). 

The top panel of Fig.~\ref{fig:mass_correlations} shows the maximum mass $m_{\rm max}$, the Hubble constant $H_0$, and their correlation to the primary mass of $\rm GW190521$, which is the most informative event for $m_{\rm max}$. As expected, the primary mass of $\rm GW190521$ is strongly positively correlated with the population mass cutoff $m_{\rm max}$. It is also mildly anti-correlated with $H_0$. This is because the source-frame mass is obtained, for given values of the cosmological parameters, from the observed detector-frame mass and luminosity distance. One has $m = m_z/[1+z(d_L, H_0,  \Omega_{\rm m})]$, where the redshift is computed by inverting the luminosity-distance redshift relation. At low redshift, this reduces to the Hubble law $z \sim H_0\, d_L$, thus $ m \sim m_D/(1+H_0\, d_L) $. Hence, for a given luminosity distance and detector-frame mass, increasing (decreasing) $H_0$ decreases (increases) the resulting source-frame mass, which explains the correlation.

Interestingly, we find a similar trend for the event $\rm GW190924\_021846$ and the parameters $m_{\rm min}$ and $\delta_{\rm m}$ shaping the lower edge of the mass spectrum --  in particular, the secondary mass of $\rm GW190924\_021846$ is positively correlated with $m_{\rm min}$ and negatively correlated with $\delta_{\rm m}$.

The middle panel of Fig.~\ref{fig:mass_correlations} shows the mean of the Gaussian component $\mu_{g}$ entering the primary mass spectrum, the Hubble constant $H_0$, and their correlation to the primary mass of $\rm GW200224\_222234$, which is the most informative event for $H_0$ and the second most informative event for $\mu_{g}$, nearly at the same level as $\rm GW190727\_06033$ (cf. Figure~\ref{fig:contributions}). Similarly, the Hubble constant and the BH mass are anti-correlated.

Finally, the bottom panel of Fig.~\ref{fig:mass_correlations} shows correlations between the primary spin of $\rm GW190517\_055101$ and the population parameters entering the spin magnitude distribution. This event has the largest effective spin in the catalog, and, consistently with this, we find that it is positively correlated to the variance $\sigma_{\chi}$ of the spin distribution.

\section{Conclusions}\label{sec:conclusions}

In this work, we have demonstrated that sampling the full hierarchical likelihood in GW population inference is not only possible but also provides some additional insights. This approach enables direct access to both individual event parameters and population properties eliminating the uncertainties associated with some of the Monte Carlo integrations involved in the likelihood evaluation. 

Modern PPLs such as {\sc PyMC} (used for this paper) provide a natural framework for formulating complex hierarchical models while allowing a variety of computational advantages, including automatic differentiation, compilation, and the automatic optimization of the likelihood computational graph. They also support the implementation of non-parametric models, which are becoming increasingly popular in our field~\cite{Edelman:2022ydv,Golomb:2022bon,Callister:2023tgi,Ray:2023upk,Sadiq:2023zee,Rinaldi:2024eep,Farah:2024xub,Heinzel:2024jlc,Fabbri:2025faf}. With the scalability of GW analysis becoming a pressing issue, these frameworks should be explored more prominently to face the imminent arrival of new data.
PPL implementations run seamlessly on GPUs, with promising large-scale applications in contexts where speed is crucial such as low-latency pipelines.

As for our population fits, sampling the full hierarchical model eliminates uncertainties related to Monte Carlo integration for single-event posteriors. At the same time, some related challenges remain untackled.
First, results remain sensitive to the number and distribution of software injections needed to compute the selection function $\xi(\lambda)$ ---which remains the only Monte Carlo integral involved in the likelihood evaluation--- and the related variance cut imposed on the likelihood, whose value remains somewhat arbitrary. 
Promising advances in this direction rely on constructing selection-function emulators using machine-learning techniques \cite{Gerosa:2020pgy,Talbot:2020oeu,Callister:2024qyq}. 
However, this might result in new, and arguably harder to quantify, sources of systematic uncertainty from the emulator performance. Other ideas that still need to be deployed in GW astronomy include adopting suitable loss functions to keep the systematic error under control and propagating the uncertainty related to the network prediction to the posterior probability distribution~\cite{arbel2023primerbayesianneuralnetworks,he2025surveyuncertaintyquantificationmethods}.
%\MM{
One further possibility is eliminating the need for this Monte Carlo integration at the root and moving towards fitting the \emph{observed} population directly. 
More drastically, modern simulation-based inference methods offer the tantalizing opportunity to remove the likelihood function altogether~\cite{Leyde:2023iof}.

Our implementation relies on density-estimating the individual-event posterior distributions from existing samples. The accuracy of the underlying Gaussian Mixture Model fit, while sufficient for the analysis presented here, might need to be improved in the future as the amount of data increases. Normalizing flows~\cite{Papamakarios2021} provide a promising way forward in this direction. Other approaches have also been explored, e.g. normal approximations~\cite{Delfavero:2022pnq}, truncated Gaussian
mixture models~\cite{Hussain:2024qzl}, or Dirichelet processes~\cite{Rinaldi:2021bhm}.

Looking further ahead, the framework discussed in this paper could also be particularly valuable in the context of dark-siren cosmology, where the population priors are informed by galaxy catalogs~\cite{Moresco:2022phi,Mastrogiovanni:2024mqc}. The full likelihood in this scenario requires the computation of integrals convolving the GW likelihood with rapidly varying ``line-of-sight'' redshift priors, currently computed with costly quadrature~\cite{Gray:2023wgj,Borghi:2023opd} or Monte Carlo methods~\cite{Mastrogiovanni:2023emh}.
These techniques could benefit from the sampling methods demonstrated here to achieve robust results and address related scientific questions such as inferring the host galaxy properties of merging compact objects~\cite{Vijaykumar:2023bgs,Perna:2024lod}.

In conclusion, we hope the framework presented in this paper can offer significant improvements in accuracy and flexibility for GW population inference in view of increasingly informative catalogs.

\acknowledgements

We thank Matthew Mould, Colm Talbot, Viola De Renzis, Alessandro Agapito, and Chiara Anselmo for discussions. %
% Michele
M.M. is supported by the French government under the France 2030 investment plan, as part of the Initiative d'Excellence d'Aix-Marseille Universit\'e -- A*MIDEX AMX-22-CEI-02.
% Davide
D.G. is supported by 
ERC Starting Grant No.~945155--GWmining, 
Cariplo Foundation Grant No.~2021-0555, 
MUR PRIN Grant No.~2022-Z9X4XS, 
MSCA Fellowships No.~101064542--StochRewind and No.~101149270--ProtoBH, 
MUR Grant ``Progetto Dipartimenti di Eccellenza 2023-2027'' (BiCoQ), 
and the ICSC National Research Centre funded by NextGenerationEU.   
Computational work was performed at CINECA with allocations through INFN and Bicocca. 
This document has LIGO DCC number LIGO-P2500055.

This research has made use of data or software obtained from the Gravitational Wave Open Science Center (gwosc.org), a service of the LIGO Scientific Collaboration, the Virgo Collaboration, and KAGRA. This material is based upon work supported by NSF's LIGO Laboratory which is a major facility fully funded by the National Science Foundation, as well as the Science and Technology Facilities Council (STFC) of the United Kingdom, the Max-Planck-Society (MPS), and the State of Niedersachsen/Germany for support of the construction of Advanced LIGO and construction and operation of the GEO600 detector. Additional support for Advanced LIGO was provided by the Australian Research Council. Virgo is funded, through the European Gravitational Observatory (EGO), by the French Centre National de Recherche Scientifique (CNRS), the Italian Istituto Nazionale di Fisica Nucleare (INFN) and the Dutch Nikhef, with contributions by institutions from Belgium, Germany, Greece, Hungary, Ireland, Japan, Monaco, Poland, Portugal, Spain. KAGRA is supported by Ministry of Education, Culture, Sports, Science and Technology (MEXT), Japan Society for the Promotion of Science (JSPS) in Japan; National Research Foundation (NRF) and Ministry of Science and ICT (MSIT) in Korea; Academia Sinica (AS) and National Science and Technology Council (NSTC) in Taiwan.

\section*{Data availability}
The code used for this work is available at~Ref.~\cite{pymcpop}. 
 This paper is associated to version v0.1.0 which is archived on Zenodo at~Ref.~\cite{zenodo_code}.
%\href{https://github.com/CosmoStatGW/pymcpop-gw}{github.com/CosmoStatGW/pymcpop-gw}, 
Data products are available at~Ref.~\cite{zenodo_release}.
\bibliography{fullhierarchical}

\appendix
\begin{figure}[b]
    \includegraphics[width=.5\textwidth]{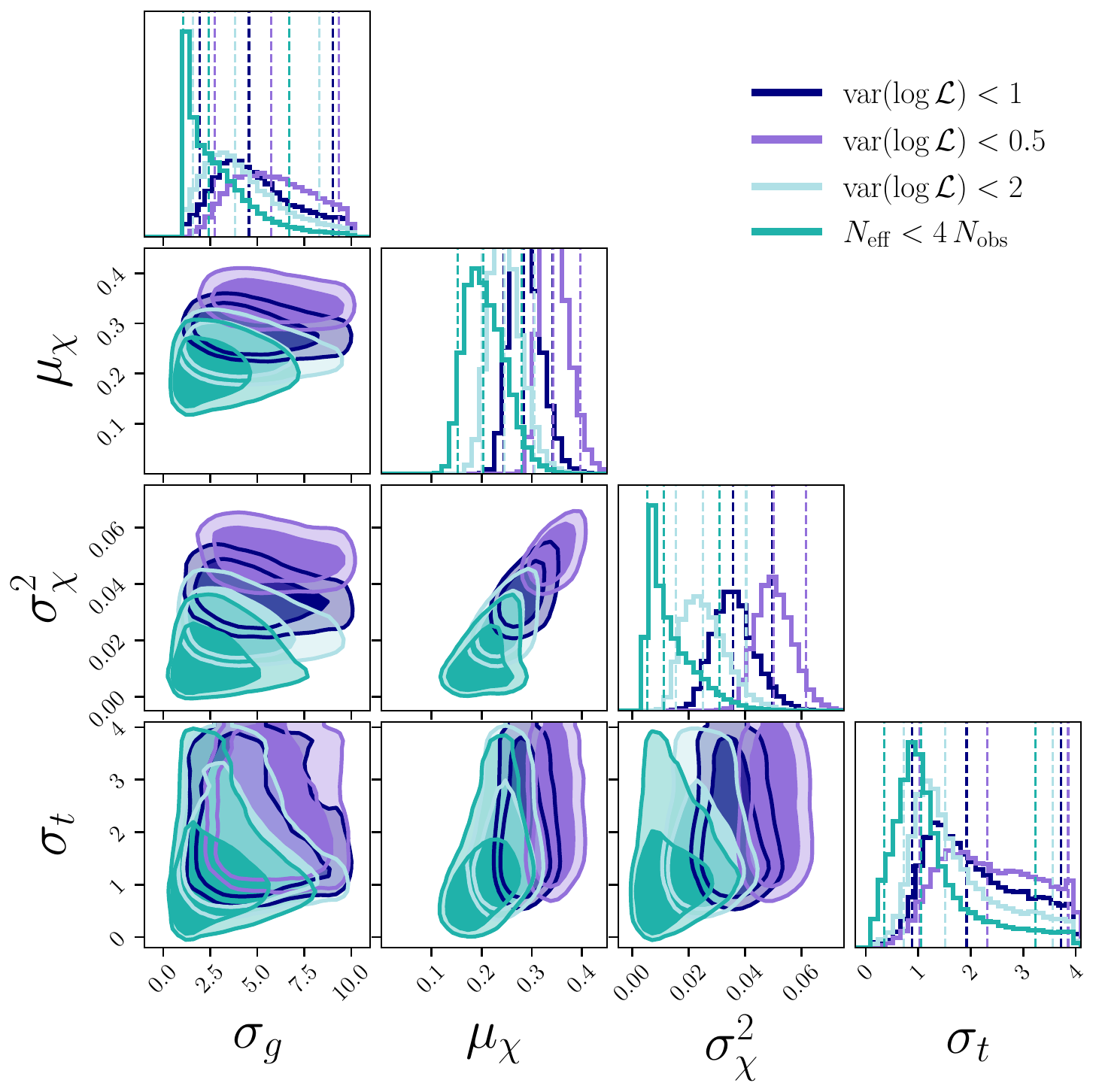}
    \caption{Comparison of inference results with different cuts on the log--likelihood variance for a subset of the population parameters. The blue distribution corresponds to the choice made in the rest of the paper, where we threshold on the likelihood variance; the purple and light blue distributions explore variations around that threshold value; the teal distribution instead considers a threshold on the effective number of events.     Contours correspond to $68\%$ and $90\%$ credible intervals.
}   \label{fig:corner_cuts}
\end{figure}

\section{Sensitivity to accuracy cuts}\label{sec:accuracy}

The variance of Monte Carlo integrals is a key source of uncertainties in hierarchical Bayesian fits. In particular, the choice of likelihood variance thresholds can impact the inference on population features, especially those that are narrower than the typical uncertainty on the underlying single--event parameters~\cite{Talbot:2023pex}. 
For the model explored in this paper, we find that this issue is particularly severe for the mean and variance $\{\mu_{\chi},\ \sigma_{\chi}\}$ of the Beta distribution parametrizing the BH spin magnitudes, for the variance $\sigma_{\rm g}$ of the Gaussian component of the primary-mass distribution and for the variance $\sigma_{\rm t}$ of the Gaussian component of the spin-orientation distribution. In Fig.~\ref{fig:corner_cuts}, we show additional results where we assume harder (light blue) or softer (purple) thresholds on the likelihood variance with respect to the standard choice of 1 (blue). For comparison, we also show results where we instead threshold using the effective number of Monte Carlo samples $N_{\rm eff}<4\times N_{\rm obs}$ (teal). While this is the same choice made in Ref.~\cite{KAGRA:2021duu} for the selection function,  the results of Fig.~\ref{fig:corner_cuts} are not directly comparable to theirs because of the additional threshold on the effective number of samples for each event, c.f. Sec.~\ref{accreq}.
Overall, these results indicate that developing robust ways to evaluate selection effects is crucial  in current GW population analyses.

\end{document}